# Precision UBVJH Single Channel Photometry
# of
# Epsilon Aurigae

*Jeffrey L. Hopkins*
*Hopkins Phoenix Observatory*
*7812 West Clayton drive*
*Phoenix, Arizona 85033-2439*
*phxjeff@hposoft.com*

*Robert E. Stencel*
*University of Denver*
*Denver, Colorado*
*rstencel@du.edu*

**Abstract**

First observed in the early 1980's the Hopkins Phoenix Observatory continues its UBV band observations of the long period (27.1 years) eclipsing binary star system epsilon Aurigae. The UBV observations routinely produce standard deviations or data spread better than 0.01 magnitudes many times approaching 0.001 magnitudes. A new infrared detector has allowed the addition of precision infrared observations for the JH bands. Typical infrared observations approach a standard deviation of data spread of 0.01 magnitudes. The 2003 - 2005 seasons (Fall through Spring) of epsilon Aurigae observations showed a 66.2 day variation that gradually increases in average and peak magnitude in the UBV bands, The 2006 season (Fall 2006 to Spring 2007) data show what appears to be a fall-back to a quiet period near maximum amplitude of V= 3.00. This paper presents the data and compares the current season to the past several seasons. The next eclipse is scheduled to begin in 2009 and an international campaign has been organized to coordinate new observations. [http://www.~hposoft.com/Campaign09.html]

## 1. Introduction

First observed by German astronomer J. Fritsch in 1821, this system has been observed in detail through most of the eclipses since. A concentrated effort was put forth during the 1982 - 1984 eclipse. While much was learned, many questions still went unanswered.

Most astronomers interested in epsilon Aurigae tend to only get excited about it as the 27.1 year eclipse approaches. For many astronomers they miss the eclipse altogether. For dedicated observers this may happen twice or even three times in a lifetime. Each time great changes in astronomy have occurred. The next eclipse will be starting in 2009. While the primary eclipse is certainly interesting, the star system also has more than a few surprises out-of-eclipse.

The Hopkins Phoenix Observatory has been observing epsilon Aurigae between the fall of 1982 and the winter of 1988 with a renewed concentrated effort beginning in the fall of 2003.

## 2. Why Observe Epsilon Aurigae?

Why all the interest in this long period eclipsing binary system? Aside from holding the record as being the longest known eclipsing binary star system, the object that eclipses the primary F supergiant seems to be a fantastically large opaque disk and has been called a round paving brick with a hole in the middle. The primary star is no small star either with an estimated diameter of 200 solar diameters. While that is certainly large it pales in comparison to the secondary, which is estimated at 2,000 solar diameters. Placed in our solar system where the Sun is located, even the orbit of Saturn would be millions of miles beneath the surface.

One of the problems with the epsilon Aurigae eclipse is the determination of contact times. This is because of the out-of-eclipse variations that skew the magnitude around contact points. By extensive out-of-eclipse observations it is hoped to better understand these variations to increase the accuracy of contact time determination as well as just what is going on in the star system.

Figure 1 shows a diagram of the system as current understood.

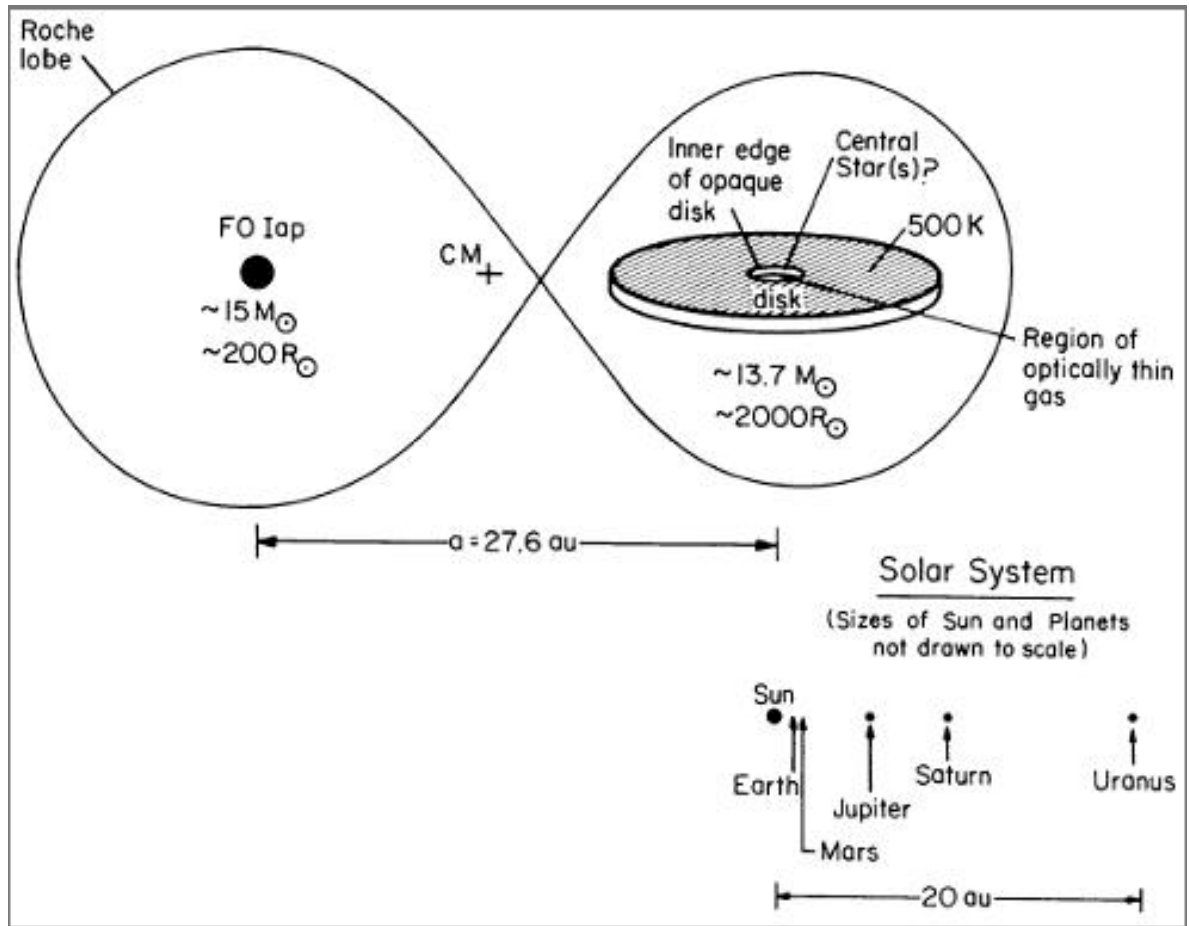

**Figure 1. The epsilon Aurigae Star System (current interpretation) Carroll et al. 1991 Ap.J. 367**

## 3. Epsilon Aurigae Questions

The main goal for observing epsilon Aurigae out-of-eclipse is to provide a baseline for analysis of data obtained during the coming eclipse. While the model presented above is generally accepted, the nature of the dark secondary is not well established. In addition, these pre-eclipse data pose questions as well:
1. What causes the out-of-eclipse variations? Some possible answers:
   a. Primary Star pulsations?
   b. Additional eclipsing material?
   c. Pulsations of the Secondary?
   d. Eclipsing of the binary pair in the secondary?
   e. Some other phenomena?
   f. Some combination of the above?
2. What can be deduced for the observed variations?
3. Why different amplitude variations in different bands?
4. What is the cause of the increasing/decreasing of average/maximum/minimum amplitudes in various bands?
5. What causes the color changes (B-V), (U-B) and (J-H)?

## 4. Equipment

As reported last year[1] the Hopkins Phoenix Observatory continues to use two photometric systems for observing epsilon Aurigae. Both systems are single channel photometers. The UBV photon counting photometer uses a HPO 1P21 photomultiplier tube based photometer with an 8" C-8 f/10 SCT. The other system is for J and H band infrared photometry. That system uses an Optec SSP-4 analog photometer with a 12" LX200 GPS f/10 SCT. The SSP-4 employs a thermoelectric cooled detector, which is cooled to around -40 degrees.

## 5. Technique

**UBV Observations**

To obtain optimum precision and accuracy a method was developed for the UBV single channel photon counting photometry that includes taking 3 sets of star readings in each band followed by one sky reading in each band. The program star (epsilon Aurigae) observations are bracketed by comparison star (lambda Aurigae) readings.

**Extinction Considerations**

Because these stars are nearly 5 degrees apart, extinction becomes important even when doing differential photometry. While extinction is minimized near the meridian and this is where the majority of observations have been made, it becomes increasingly important to consider an accurate extinction the further the stars are from the meridian.

Previously, average extinction coefficients were used. To increase the accuracy it was decided to determine nightly extinction. This was done using the comparison star's nightly readings along with the star's air mass to calculate nightly K'v, K'bv and K'ub. While the use of nightly extinction coefficients as opposed to average coefficients had a minor effect on the magnitudes it proved important when tracking magnitudes in the 0.001 to 0.01 magnitude region.

Figure 2 shows a screen shot of a FileMaker Pro database program developed by HPO for UBV band photometry. This screen shot shows a summary of a night's data including Averaged Net Star Counts, HJDs and Average HJD, Air Masses and calculated Observation Extinction Coefficients.

**Figure 2. Typical UBV Data Entry Summary Screen**

**JH Band Infrared Observations**

The SSP-4 J and H band photometry procedure is a bit different from the UBV procedure. Epsilon and lambda Aurigae are not bright in the infrared and even with a 12" telescope the net star counts are not very high. This requires use of the SSP-4's maximum gain and 10 second integration/gate times. One big problem with the SSP-4 photometer is that when using the maximum gain of 100 and 10-second gate time, the dark counts drift considerably, sometimes over 1,000 counts in a 30-minute time span. While the variations tend to cancel out with three sets of measurements, to maximize precision and accuracy it was found that it was best to use 3 - 10 second readings of the star + sky in each band followed by single 10 second measurements of the sky in each band. To monitor the dark counts a 10 second reading is taken at the start and end of the session and between switching of the stars. Subtracting the closest dark count from the sky + dark count gives a net sky count and allows checking of the sky.

**Note:** The sky + dark is used when determining the net star count.

One nice feature about JH band photometry is extinction is near zero. In fact, observations can be done during close full Moon, in twilight and even daylight if the stars can be accurately found.

Figure 3 shows a screen shot of a FileMaker Pro database program developed by HPO for JH band photometry. This screen shot shows a summary of a night's data including Averaged Net Star Counts, HJDs and Average HJD, Air Masses, Dark Counts and Net Sky Counts.

**Figure 3. JH Data Entry Summary Screen**

## 6. Reduced Data

Table 1 shows a list of sample reduced UBV data. Notice the low standard deviation (SD) data spreads

| HJD 2,450,000+ | V | SD | B | SD | U | SD | X |
|---|---|---|---|---|---|---|---|
| 4123.6963 | 3.016 | .000 | 3.585 | .004 | 3.700 | .007 | 1.019 |
| 4128.6470 | 2.997 | .002 | 3.564 | .002 | 3.688 | .007 | 1.021 |
| 4133.7414 | 3.005 | .002 | 3.583 | .002 | 3.704 | .013 | 1.119 |
| 4135.6067 | 3.007 | .003 | 3.576 | .008 | 3.702 | .011 | 1.037 |

**Table 1. Sample UBV Data**

Table 2 shows a list of sample reduced JH infrared data.

| HJD 2,450,000+ | J | SD | H | SD | X |
|---|---|---|---|---|---|
| 126.7086 | 1.781 | .009 | 1.438 | .013 | 1.0393 |
| 127.6662 | 1.770 | .005 | 1.448 | .008 | 1.0165 |
| 128.6634 | 1.799 | .006 | 1.437 | .015 | 1.0165 |
| 135.6224 | 1.773 | .007 | 1.413 | .011 | 1.0205 |

**Table 2. Sample JH Data**

## 7. Light Curve Analysis

Figures 4 and 5 show time domain plots of the V and data for 2003 - 2007 and 2006 - 2007 respectively.

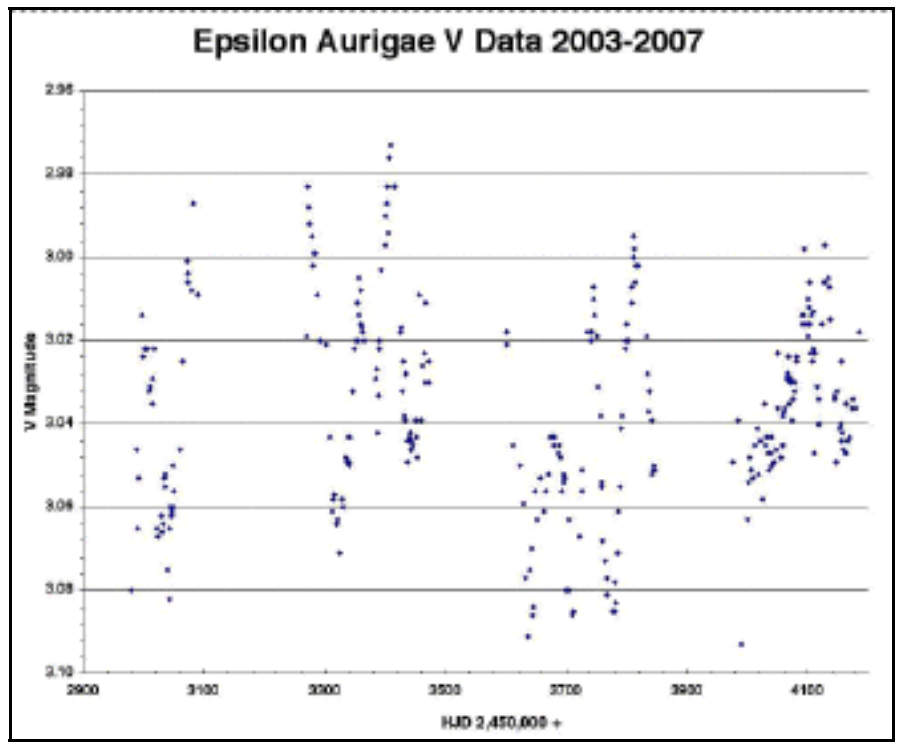
**Figure 4. V Band Light Curve(Time Domain) 2003 –2007**

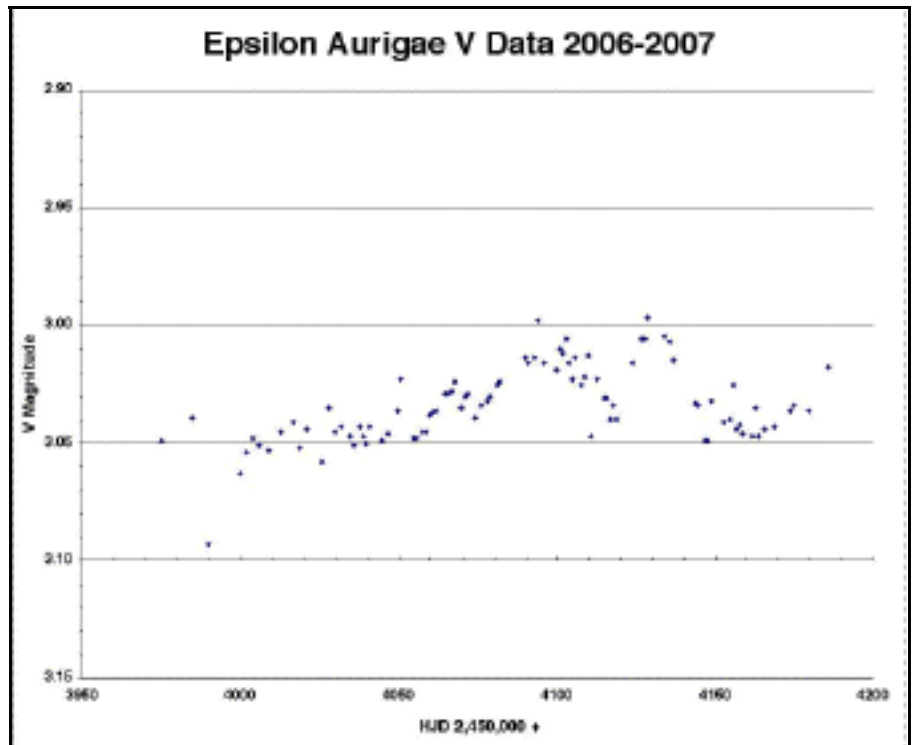
**Figure 5. V Band Light Curve(Time Domain) 2006 -2007**

Figure 6 shows a time domain plot of the J band for 2005 - 2007. The 2995/2006 season shows considerable scatter. This was due to the instability of the smaller infrared detector in the SSP-4. The 0.3 mm detector was replaced with the large 1.0 mm and the 2006/2007 season shows more consistent data.

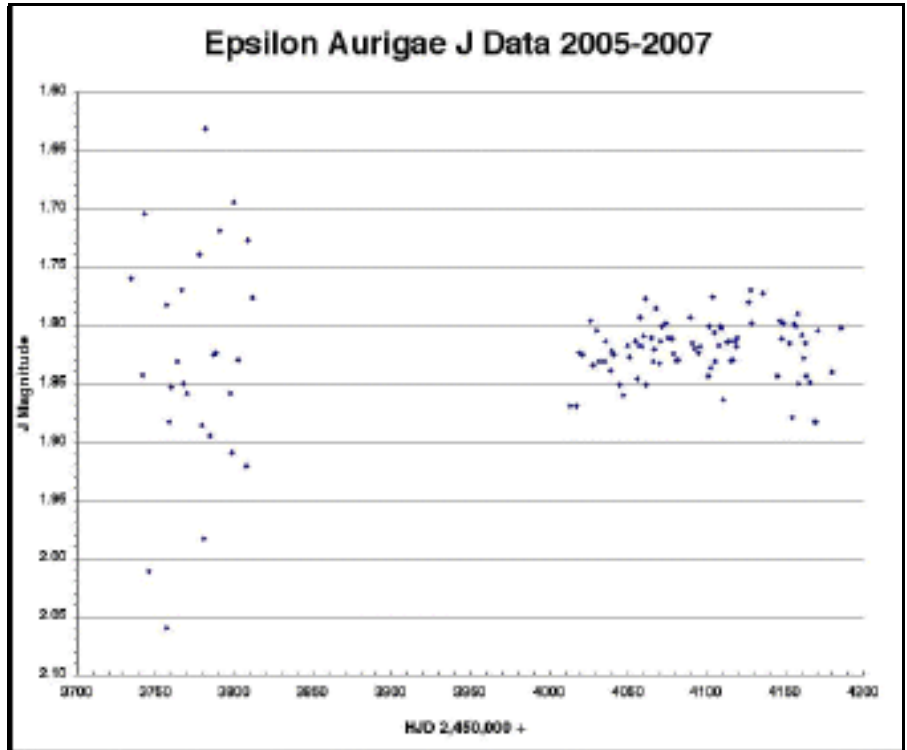
**Figure 6. J Band Light Curve(Time Domain) 2006 -2007**

## 8. Period Analysis

Frequency domain plots were obtained using Peranso software and the Discrete Fourier Transform (DFT Deeming) technique. Other techniques were tried with similar results.

Figure 7 shows a frequency domain plot of the U Band data for the 2003 - 2007 period. The main frequency peak is at 164.7604 days. The second largest peak is at 66.3964 days. The 164 day peak is due to the seasonal observations (i.e., approximated start of each season is ~164 days apart).

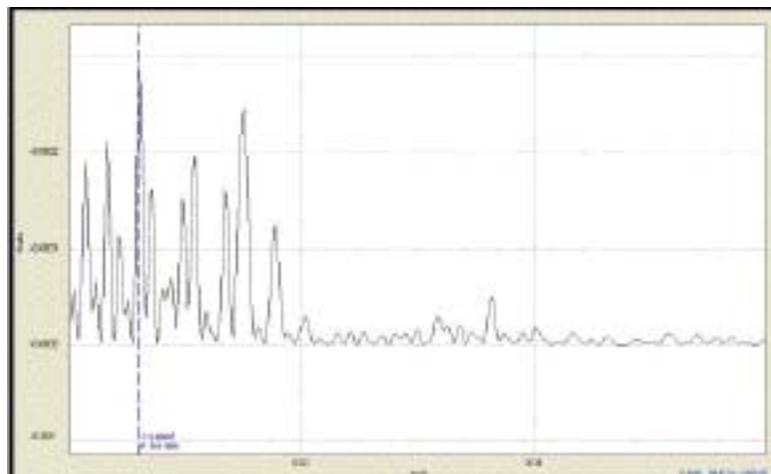
**Figure 7. U Band Frequency Domain Plot 2003 - 2007, Cursor at 164.7604 Days**

Figure 8 shows a frequency domain plot of the B Band data for the 2003 - 2007 period. The V band plot is very similar. The main frequency peak is at 66.4986 days. The second largest peak is at 167.1359 days. Again the 167-day peak is due to the period of the observation seasons.

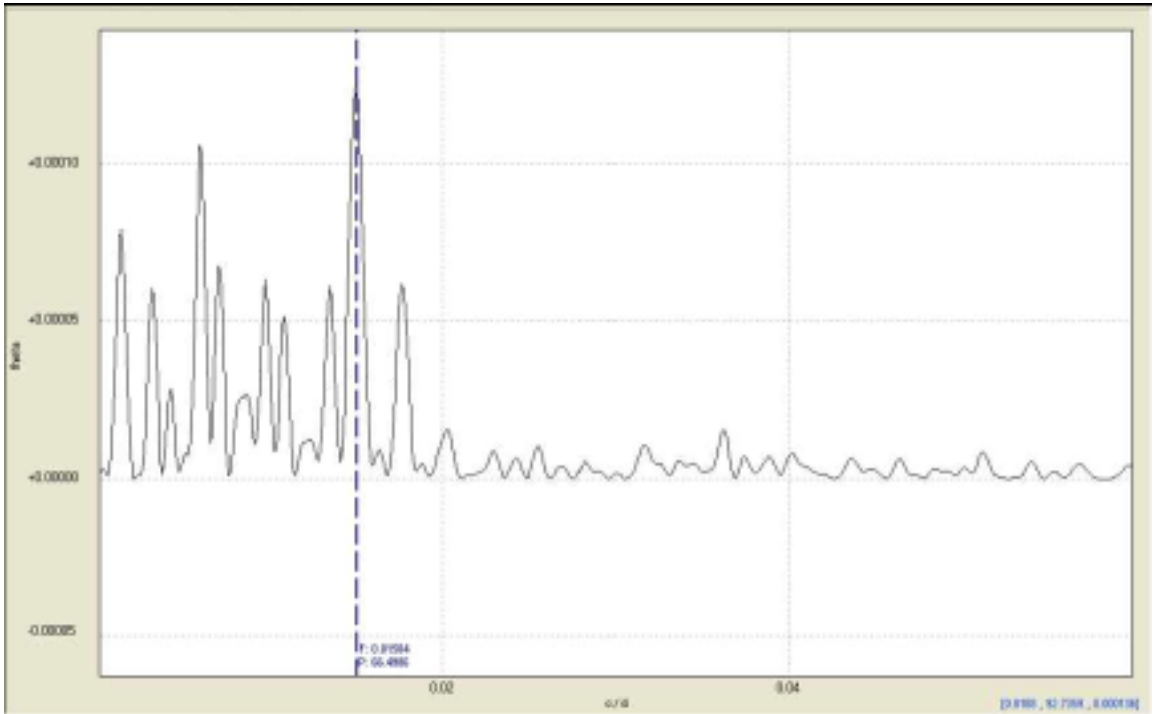

**Figure 8**
**B Band Frequency Domain Plot 2003 - 2007, Cursor at 66.4986 Days**

The J and H band data was not sufficient to produce meaningful frequency domain plots. It is hoped the next two seasons will provide ample data for a detail period analysis in the infrared bands. Table 3 lists a magnitude summary of the UBV data for the 2003 - 2007 period. The average and mean values are simple average and mean and are not weighted. Differences between the average and mean are very small and may be more of an indication of measurement error than a real difference.

| Filter | Max | Min | Delta | Average | Mean |
|---|---|---|---|---|---|
| V | 2.9726 | 3.0927 | 0.1201 | 3.0358 | 3.0327 |
| B | 3.5246 | 3.6869 | 0.1623 | 3.6051 | 3.6058 |
| U | 3.6000 | 3.9207 | 0.3207 | 3.7275 | 3.7603 |
| B-V | 0.5277 | 0.6073 | 0.0796 | 0.5693 | 0.5675 |
| U-B | 0.0576 | 0.2687 | 0.2111 | 0.1224 | 0.1631 |

**Table 3. UBV Magnitude Data Summary 2003 –2007**

Table 4 lists a period summary of the UBV data for the 2003 - 2007 period.

| Filter | Pri Period | Sec Period |
|---|---|---|
| V | 66.4986 days | 761.7669 days |
| B | 66.4986 days | 167.1359 days |
| U | 164.7604 days | 66.3964 days |
| B-V | 164.7604 days | 295.1177 days |
| U-B | 423.2912 days | 289.5742 days |

**Table 4 UBV Period Data Summary 2003 -2007**

Table 5 lists a magnitude summary of the UBV data for the 2006 - 2007 period.

| Filter | Max | Min | Delta | Average | Mean |
|---|---|---|---|---|---|
| V | 2.9967 | 3.0927 | 0.0960 | 3.0340 | 3.0447 |
| B | 3.5434 | 3.6328 | 0.0894 | 3.5984 | 3.5881 |
| U | 3.6321 | 3.7719 | 0.1398 | 3.7095 | 3.7020 |
| B-V | 0.5277 | 0.5940 | 0.0663 | 0.5644 | 0.5608 |
| U-B | 0.0647 | 0.1555 | 0.0908 | 0.1111 | 0.1101 |

**Table 5. UBV Data Summary 2006 -2007**

Table 6 lists a period summary of the JH data for the 2005 - 2007 period. A note of caution is in order. Because the 2005 to 2006 season used the smaller detector and the data was not consistent, a reliable set of periods for the JH bands will require at least one more season.

| Filter | Pri Period | Sec Period |
|---|---|---|
| J | 141.8641 days | 69.3131 days |
| H | 145.1337 days | 104.1014 days |
| J-H | 150.3307 days | 105.9853 days |

**Table 6. JH Period Data Summary 2005 -2007**

## 9. Spectroscopy Support

H-alpha spectra are being regularly acquired by Lothar Schanne[2] in Germany to provide a road map of the nebular emission from the binary system. This enables another dimension to be added to the revelations of the UBVJH photometry reported here. The line center velocity gives a Doppler measurement of the densest material, while the emission bump variation informs us of the movement of lower density clouds associated with one of the components. Once again, having this pre-eclipse record will help place in-eclipse variation into a useful context that largely was absent during the run up to the 1982 eclipse.

## 10. Conclusions

We have clearly demonstrated short term, period-like variations in the UBV light curves of epsilon Aurigae. New photometry at J and H bands is underway, but signal to noise does not presently allow us to confirm similar variations. We also see strong evidence for a persistent 66.5 day period of UBV light variation with an amplitude of ~0.1 magnitude

We find from our photometry, conducted 2003-2007 [MJD52978 - 54185], that B-V = 0.569 +/- 0.012, and U-B = 0.123 +/- 0.025. The B-V color of epsilon Aurigae is not that of a normal F0 supergiant, while U-B is more nearly so. This discrepancy is not consistent with normal interstellar reddening.

Gyldenkerne (1970) reported on photometry obtained during the 1958 eclipse with a 10-inch telescope and EMI5060 photomultiplier, and transformed using calibration stars to the Johnson B and V filters. He stated the mean pre-eclipse values were V = 2.998 (N = 50 points) and B-V = 0.555 (38 points); mid-eclipse values were V = 3.767 (117), B-V = 0.575 (62), and post-eclipse values returned to V = 3.006 (31) and B-V = 0.554 (37). Furthermore the detailed a correlation between delta V and delta B-V during totality, described by the equation $\Delta(B-V) = 0.320 * \Delta V - 0.005$, where $\Delta V$ spanned +/- 0.1 magnitude and $\Delta(B-V)$ spanned +/- 0.05 magnitude. We see evidence for similar behavior in the present data.

Trends in the UBV light curve are suggestive that the outer edges of the opaque disk may be participating in an increasing amount of forward scattering of F starlight toward the observer. If the trends persist and increase during the 2007/08 and 2008/09 observing seasons, we can include those effects in more comprehensive analysis of the eclipse phenomena scheduled to start in 2009 August.

Figure 9 shows a plot of (U - B) vs. (B - V).

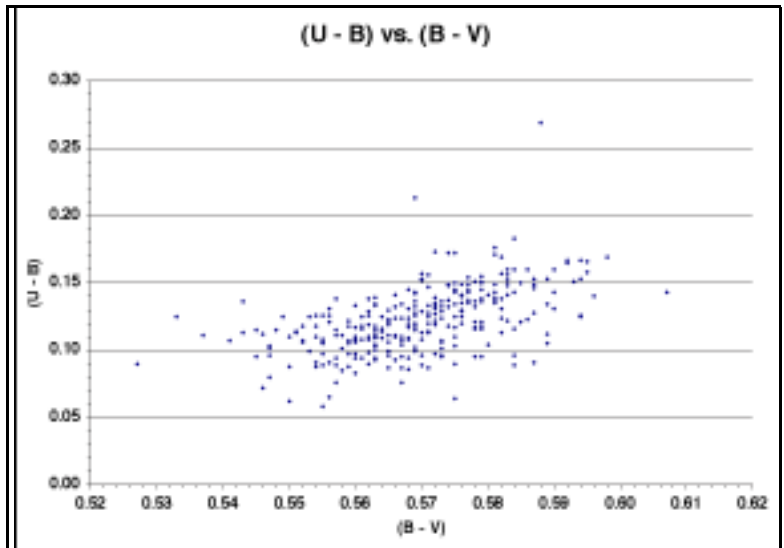
**Figure 9. (U - B) vs. (B - V)**

Figure 10 shows a schematic of the star system indicating configurations before and during the eclipse.

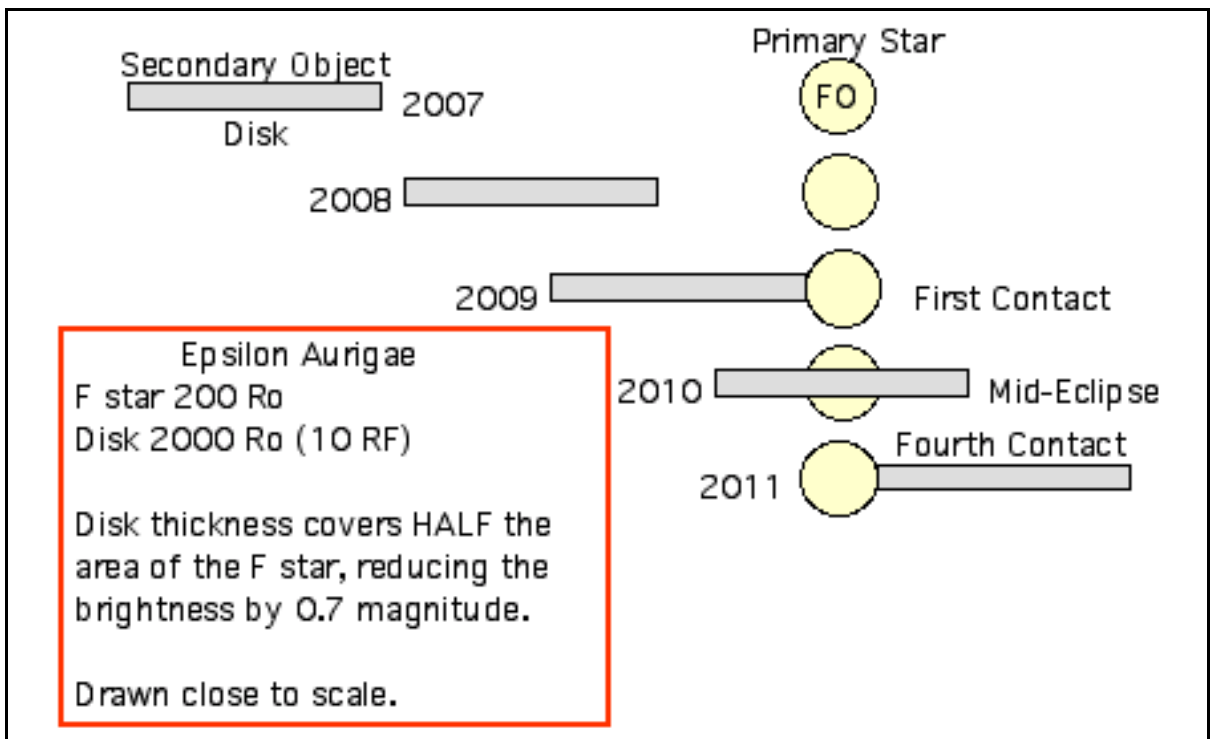
**Figure 10. Epsilon Aurigae Eclipse Timing Schematic**

## 11. Epsilon Aurigae Campaign

The next eclipse of epsilon Aurigae is fast approaching. As mentioned last year an eclipse campaign has been started with and a web site has been developed to disseminate information and coordinate efforts for the next eclipse and for events prior to the eclipse. The **Epsilon Aurigae Campaign 2009** web site can be accessed at:

http://www.hposoft.com/Campaign09.html

Those interested in the eclipse are encouraged to check the web site and join the campaign. In particular we are interested in confirmation of short and intermediate variations in all photometric bands and an extension of coverage during the summer months (the higher the observers latitude the better).

## 12. Other Similar Star Systems

There are a couple of other stars systems similar to epsilon Aurigae that bear observing. One is the 5.61-year eclipsing binary system EE Cephei. The next eclipse starts in January 0f 2009 and lasts 17 days. See references at the end of this paper for additional information. Additional information on EE Cephei can be found at:

http://www.hposoft.com/EAur09/EECephei.html

Another system is the short period (6.5 days) eclipsing binary system BM Orionis. BM Orionis is of interest because the star system produces an eclipse very similar to that of epsilon Aurigae, but much shorter (period of only 6.5 days versus 27 years for epsilon Aurigae). Like the eclipse of epsilon Aurigae the BM Orionis eclipse is flat-bottomed and even has a mid-eclipse brightening. Like the secondary of the epsilon Aurigae system the BM Orionis secondary appears to be a brick with a hole in it. This is an extremely difficult star system to do single channel photometry on, however, CCD photometry is easier, but still challenging. Additional information on EE Cephei can be found at:

http://www.hposoft.com/EAur09/BMOrionis.html

## 13. References

1- Proceedings for the 25th Annual Conference of the Society for Astronomical Sciences, May 2006, *Single Channel UBV and JH Band Photometry of Epsilon Aurigae*, Jeffrey L. Hopkins, Hopkins Phoenix Observatory, 7812 West Clayton Drive, Phoenix, Arizona 85033-2439 and Robert E. Stencel, University of Denver, Denver, Colorado

2. IBVS 5747, *Remarkable Absorption Strength Variability of the epsilon Aurigae H alpha Line outside Eclipse*, 9 January 2007, Schanne, L., Hohlstrasse 19, D-66333 Völklingen (Germany)

**EE Cephei References**

IBVS 1225, *VARIABLE STAR IN EE CEPHEI COMPARISON SEQUENCE*, L. BALDINELLI, S. GHEDINI, "G. Horn D'Arturo" Observatory, Montebello 4, 40121 Bologna, Italy, 30 December 1976

IBVS 1939, *1980 Eclipse of EE Cephei: Light Curve and Time of Minimum*, 23 March 1981, L. Baldinelli, A. Ferri, S. Ghedint, "G. Horn D'Arturo" Observatory, C.P. 1630 AD, 40100 Bologna, Italy

IBVS 5412, *The start of the 2003 eclipse of EE Cephei*, Mikolajewski, M.1; Tomov, T.1; Graczyk, D.1; Kolev, D. 2; Galan, C.1; Galazutdinov, G.1,3

*The 2003 eclipse of EE Cep is coming, A review of past eclipses.* D. Graczyk1, M., Miko•ajewski1, T. Tomov1, D. Kolev2, and I. Iliev2
17 March 2003, Astronomy & Astrophysics

*The 2003 Eclipse of EE Cephei,* Gerard Samolyk, 9504 W Barnard Ave, Greenfield WI 53228. Shawn Dvorak, 1643 Nightfall Drive, Clermont FL 34711, JAAVSO Volume 33, 2004

*Timing of the 1997 Eclipse of the Long Period (5.61 Years) Eclipsing Binary EE Cephei*, Edward A. Halbach, JAAVSO Volume 27, 1999 (.pdf of article is available at
http://www.hposoft.com/EAur09/EE%20Cep/EECep99.pdf